\documentclass[useAMS,usenatbib]{mn2e}
\usepackage{graphicx}
\usepackage{amsmath}
\usepackage{amssymb}
\usepackage{color}
\usepackage{url}

\newcommand  \acc     {\ifmmode {\rm km\,s}^{-2} \else km\,s$^{-2}$\fi}

\newcommand  \ergs     {\ifmmode {\rm ergs\,s}^{-1} \else ergs s$^{-1}$\fi}
\newcommand  \ergcms   {\ifmmode {\rm erg~cm}^{-2}\,{\rm s}^{-1}
                        \else erg~cm$^{-2}$\,s$^{-1}$\fi}
\newcommand  \ergcmsA  {\ifmmode{\rm erg\,cm}^{-2}\,{\rm s}^{-1}\,{\rm\AA}^{-1}
                        \else ergs\,cm$^{-2}$\,s$^{-1}$\,\AA$^{-1}$\fi}
\newcommand  \ergcmsHz {\ifmmode{\rm ergs\,cm}^{-2}\,{\rm s}^{-1}\,{\rm Hz}^{-1}
                        \else ergs\,cm$^{-2}$\,s$^{-1}$\,Hz$^{-1}$\fi}
\newcommand  \phcms    {\ifmmode {\rm ph\,cm}^{-2}\,{\rm s}^{-1}
                        \else ph\,cm$^{-2}$\,s$^{-1}$\fi}
\newcommand  \phcmsA   {\ifmmode {\rm ph\,cm}^{-2}\,{\rm s}^{-1}\,{\rm\AA}^{-1}
                        \else ph\,cm$^{-2}$\,s$^{-1}$\,\AA$^{-1}$\fi}

\newcommand\aj{{AJ}}%
\newcommand\araa{{ARA\&A}}%
\newcommand\apj{{ApJ}}%
\newcommand\apjl{{ApJ}}%
\newcommand\apjs{{ApJS}}%
\newcommand\aap{{A\&A}}%
\newcommand\aaps{{A\&AS}}%
\newcommand\mnras{{MNRAS}}%
\newcommand\pasp{{PASP}}%
\newcommand\pasj{{PASJ}}%
\newcommand\nat{{Nature}}%
\title[SN Delay Time in the Magellanic Clouds]{
The Supernova Rate and Delay Time Distribution in the Magellanic Clouds}

\author[Maoz \& Badenes] {Dan Maoz$^{1}$\thanks{E-mail: maoz@astro.tau.ac.il},
  Carles Badenes$^{1,2}$\\
  $^{1}$School of Physics and Astronomy, Tel-Aviv University, Tel-Aviv 69978,
  Israel\\
  $^{2}$Benoziyo Center for Astrophysics, Weizmann Institure of Science, Rehovot 76100, Israel\\
} \date{\today}

\begin{document}

\maketitle

\label{firstpage}

\begin{abstract}
  We use the supernova remnants (SNRs) in the two Magellanic Clouds
  (MCs) as a supernova (SN) survey, ``conducted'' over tens of 
kyr, from which we derive the current SN rate, and the 
SN delay time distribution (DTD), i.e., the SN rate vs. time that would follow a
  hypothetical brief burst of a star formation. 
In a  companion paper (Badenes, Maoz, \& Draine 2010) we have 
compiled a list of 77 SNRs in the MCs, and argued that it
is a fairly complete record of the SNRs that are now in the
Sedov phase of their expansions.  
 We recover the SN DTD by comparing the
  numbers of SNRs observed in  small individual ``cells'' in these galaxies
 to the star-formation histories of each cell,
 as calculated from resolved stellar populations by Harris \&
  Zaritsky. We identify the 
visibility times of SNRs in each cell with the Sedov-phase lifetimes, 
which depend on the local ambient densities.
The local densities
 are estimated from 21-cm emission, from an inverse
  Schmidt-Kennicutt law based on either H$\alpha$ emission or on the 
star-formation rate from the resolved stellar populations, and from 
combinations of these tracers. This is the first SN DTD that is based
on resolved stellar populations. We detect a population
of ``prompt'' type-Ia SNe (that explode within 330~Myr of star
  formation) at $>99\%$ confidence level (c.l.). The best
  fit for the number of prompt type-Ia SNe  per stellar mass formed
is $2.7-11.0\times 10^{-3}M_\odot^{-1}$, depending on the density
  tracer used. The $95\%$ c.l. range for a ``delayed'' (from
  $330$~Myr to a Hubble time) type-Ia component is $<1.6\times 10^{-13}$~SN~yr$^{-1}M_\odot^{-1}$,
 consistent with rate measurements in old populations. The
  current total (core-collapse+Ia) SN rate in
  the MCs is 2.5-4.6 SNe per millenium (68\% c.l.+systematics), 
or 1.7-3.1 SNuM [SNe $(100~{\rm
  yr}~10^{10}M_\odot)^{-1}$], in agreement with the
  historical record and with rates measured in other dwarf irregulars. Conversely,
  assuming the SNRs are in free-expanion, rather than in their Sedov
  phase, would impose on the SNRs a maximum age of
6~kyr, and would imply a MC SN rate per unit mass that is 
5 times higher than in any type of galaxy, and a low-mass limit 
for core-collapse progenitors in conflict with stellar
  evolution theory. 

\end{abstract}


\section{Introduction}

Supernova (SN) explosions and their remnants touch upon multiple aspects of astrophysics and cosmology, whether as
endpoints in stellar evolution, as sources of energy and enriched material to the interstellar and intergalactic media,
as sites of cosmic-ray acceleration, or as standard candles for cosmography. However, much remains to be understood
regarding these events, both the core-collapse (CC) SN explosions that are thought to end the lives of some or all
massive ($\ga 8M_\odot$) stars; and the type-Ia SNe (SNe~Ia), that are believed to be the thermonuclear combustions of
CO white dwarfs (WDs) that have approached the Chandrasekhar mass by accreting material from, or merging with, a
companion star.

The lower limit on the initial mass of stars that eventually undergo CC is poorly known, both theoretically
\citep{poelarends08:SuperAGB_SNe} and observationally \citep[from identification of progenitors in pre-explosion images;
e.g.][]{smartt09:CCSN_progenitors}, and could be as low as 7 $M_\odot$ or as high as 12 $M_\odot$.  The upper mass limit
on the intial mass that leads to a SN explosion, rather than (perhaps) to an explosion-less collapse into a black hole,
is also uncertain theoretically \citep{heger03:CCSNe} and observationally \citep{kochanek08:survey_nothing}. Among the
CC-SN progenitors, it is still not clear which mass range leads to which SN subtypes, such as IIP, IIn, Ib, and Ic
\citep{smartt09:CCSN_progenitors}.  Parameters other than mass -- binarity, rotation, metallicity -- likely also play an
important role in determining CC-SN type \citep[e.g.][]{eldridge08:CCSN_Progenitors_Binaries}.

For SNe~Ia, our ignorance is even greater. The progenitor systems of these events are still unknown, with two distinct
models generally considered: the single-degenerate (SD) scenario in which the WD accretes material from a normal star
\citep{whelan73:SNI_SD,nomoto82:accretingWD}, and the double-degenerate (DD) model, where the WD merges with another WD
\citep{iben84:typeIsn,webbink84:DDWD_Ia_progenitors}.  Both models suffer from theoretical and observational
problems. For SD scenarios, it is unclear how the material accreted from the donor can burn quietly on the WD surface
until the required mass (close to the Chandrasekhar limit) is reached \citep{cassisi98:H-accreting-CO-WD}. The narrow H
or He lines that are expected in nebular SN spectra for SD progenitors have never been found in normal Ia events
\citep{leonard07:H_nebular_Ia_spectra}, nor have the integrated X-ray
emission from a population of accreting WDs undergoing slow nuclear
burning on their surfaces \citep{distefano10:sss_Ia_progenitors,gilfanov10:Ia_progenitors}.
The claimed identification of the surviving donor star in the nearby Type Ia
supernova remnant (SNR) Tycho \citep{ruiz-lapuente04:Tycho_Binary,hernandez09:Tycho_G} has been recently questioned by
\citet{kerzendorf09:Tycho_G}. In the case of DD systems, there are concerns regarding the fate of merging WDs, which
might lead to accretion induced collapse instead of a SN Ia explosion \citep{saio85:DD_Mergers}. It is also unknown
whether there are enough DD progenitor candidates in the Milky Way to produce the observed Ia SN rate
\citep{napiwotzki04:SPY_04,nelemans05:SPY_IV}, although current searches might clarify this point in the near future
\citep{badenes09:SWARMS_I}.

Major progress in resolving these questions could come from knowledge of the elapsed times between the formation of a
stellar population and the explosion of some of its members as different types of SNe. Indeed, a major objective of SN
studies has been the recovery of the so-called delay-time distribution (DTD). The DTD is the SN rate as a function of
time that would be observed following a $\delta$-function burst of star formation. (In other contexts, the DTD would be
called the delay function, the transfer function, or Green's function). Knowledge of the DTD would be useful for
understanding the route along which cosmic metal enrichment and energy input by SNe proceed, but no less important, for
obtaining clues about the SN progenitor systems. Different progenitor stars, binary systems, and binary evolution
scenarios predict different DTDs. The ``border'' in progenitor masses between SNe~Ia and CC-SNe will lead to a time
border in the DTD, dictated by stellar evolution, between the two types. The age corresponding to a particular mass
border will depend on metallicity. 
For example, an $8M_\odot$, $0.4 Z_\odot$ (where $Z_\odot$ is the solar metallicity)
star lives for 41 Myr.
  For reference here and
in the rest of this paper,
Table~\ref{lifetimes} lists the
\citep{girardi00:Low_Intermediate_Mass_Stars}
lifetimes of stars of various zero-age
main sequence masses, for various metallicities. 
The mean metallicities of the Small and
Large Magellanic Clouds, 
 the two galaxies that will be at the focus of this paper,
are $0.1 Z_\odot$ and $0.4 Z_\odot$, respectively (Russell \& Dopita 1992).  

\begin{table}
\label{lifetimes}
 \centering
 \begin{minipage}{100mm}
  \caption{Stellar Lifetimes}
  \begin{tabular*}{80mm}{@{}crrr@{}}
    \hline
   Mass  &  \multicolumn{3}{r}{Lifetime\footnote{Girardi et al. 2000} [Myr]} \\
   $[M_\odot]$ & $0.05Z_\odot$ & $0.4Z_\odot$ & $Z_\odot$\\
    \hline
    3&318 &363 & 477 \\
    4&167 &182 & 214 \\
    5&105 &112& 121 \\
    6&74 &76 &78 \\
    7&55 &56& 55 \\
    8&40 &41&40 \\
    9&31 &33&31 \\
    \hline 
  \end{tabular*}
\end{minipage}
\end{table}

The shape of the DTD is especially important to constrain the hotly debated nature of SN~Ia
progenitors. For both of the currently popular progenitor scenarios, SD and DD, calculations of the DTD depend on a
series of assumptions regarding initial conditions (initial mass function -- IMF , binarity fraction, mass ratio
distribution, separation distribution), and complex physics (mass loss, mass transfer, common envelope evolution,
accretion) that are often computationally intractable except in the most rudimentary, parametrized forms
\citep[e.g.,][Bogomazov \& Tutukov 2009; Mennekens et al. 2010]{yungelson98:SN_Ia_progenitors,hurley02:binaries_tides,han04:SDchannel_for_SNIa,nelemans05:common_envelope_in_WD_binaries,beer07:common_envelope,ruiter09:SNIa_rates_delay_times,bear10:common_envelope}. Observational
estimates of the DTD could rule out particular DTD predictions, or at least could provide some input and generic
features that successful models will need to reproduce.

However, to date, observational SN DTD estimates have been few and controversial.  One approach has been to compare the
SN rate in field galaxies, as a function of redshift, to the cosmic star formation history (SFH). Given that the DTD is
the SN ``response'' to a short burst of star formation, the SN rate versus cosmic time will be the convolution of the
full SFH with the DTD. \citet{gal-yam04:redshift_Ias} carried out the
first such comparison, using a small sample of SNe Ia out
to $z=0.8$, and concluded that the results were strongly dependent on the poorly known cosmic SFH, a conclusion echoed
by \citet{forster06:SNIa_progenitor_delays}.
With the availability of SN rate measurements to higher redshifts, \citet{barris04:Ia_distances} found a SN~Ia rate that
closely tracks the SFH out to $z\sim 1$, and concluded that the DTD must be concentrated at short delays, $\lesssim
1$~Gyr. Similar conclusions have been reached, at least out to $z\sim 0.7$, by \citet{sullivan06:SNIa_rate_host_SFR}. In
contrast, \citet{dahlen04:high_z_SN_rates,dahlen08:extended_HST_survey} and \citet{strolger04:higher_z} have argued for
a DTD that is peaked at a delay of $\sim 3$~Gyr, with little power at short delays, based on a decrease in the SN~Ia
rate at $z>1$. However, \citet{kuznetsova08:reanalysis_high_z_SN_rates} have re-analyzed some of these datasets and
concluded that the small numbers of SNe and their potential classification errors preclude reaching a
conclusion. Similarly, \citet{poznanski07:SNIa_rate_subaru_df} performed new measurements of the $z>1$ SN~Ia rate, and
found that, within uncertainties, the SN rate could be tracking the SFH. This, again, would imply a short delay
time. \citet{greggio08:SN_Ia_rates} pointed out that these results could also be affected by an underestimated
extinction for the highest-$z$ SNe, which are observed in their rest-frame ultraviolet emission.
  
A second approach to recovering the DTD has been to compare the SN
rates in 
stellar populations of different
characteristic ages. Using this approach,
\citet{mannucci05:SN_rate,mannucci06:two_progenitor_populations_SNIa}, \citet{scannapieco05:A+B_models},
\citet{sullivan06:SNIa_rate_host_SFR}, and \citet{raskin09:prompt_Ia_SNe} 
have 
all found evidence for the co-existence of two SN~Ia populations, a ``prompt''
population\footnote{We note
  that different authors have used the term ``prompt'' to describe
  different delays, from $<100$~Myr to as long as $< 1$~Gyr. In this
  work, we will refer as prompt SNe Ia to those with delays $<330$~Myr.} that explodes within of order  $10^8$~yr, and a delayed channel that produces SNe~Ia on timescales of order
5~Gyr. Naturally, these two ``channels'' may in reality be just integrals over a continuous DTD on two sides of some
time border \citep{greggio08:SN_Ia_rates}. \citet{totani08:SNIa_DTD} have used a similar approach to recover the DTD, by
comparing SN~Ia rates in early-type galaxies of different characteristic ages. They find a DTD consistent with a
$t^{-1}$ form. Additional recent studies using this approach, but which
may be influenced by selection effects and by \textit{a posteriori} statistics \citep[because they focus on the
properties of SN host galaxies; see][]{maoz08:fraction_intermediate_stars_Ia_progenitors} can be found in
\citet{aubourg07:massive_stars_SNIa},
\citet{cooper09:metallicity_bias_SNIa}, 
Schawinski (2009), and \citet{yasuda10:SNIa_luminosity}.

A third approach for recovering the DTD is to measure the SN-rate vs. redshift in massive galaxy clusters.  Optical
spectroscopy and muliwavelength photometry of cluster galaxies has shown that the bulk of their stars were formed within
short episodes at $z\sim 2-3$ \citep[e.g.][]{eisenhardt08:clusters}. Thus, the observed SN rate vs. cosmic time (since
the stellar formation epoch) essentially provides a direct measurement of the form of the DTD.  Furthermore, the record
of metals stored in the intracluster medium (ICM) constrains the number of SNe that have exploded, and hence the
normalization of the DTD. \citet{maoz04:Ia_rate_clusters} analysed the then-available cluster SN~Ia rates
\citep{gal-yam02:HST_SNIa_cluster_rates}.  The low observed SN~Ia rates out to $z\sim 1$ implied that the large number
of events needed to produce the bulk of the iron occurred at even higher redshifts, beyond the range of the
then-existing observations. They concluded that most of the cluster
SNe~Ia 
exploded during the relatively brief time interval between star formation in massive clusters
(at $z\sim 2-3$) and the highest-redshift cluster SN rate measurements (at $z\sim 1$).  Cluster SN rate measurements
have by now greatly improved in the redshift range from zero to 1, and beyond
\citep[][]{sharon07:SNIa_cluster_rate,gal-yam08:SN_rates_low_z_clusters,mannucci08:SN_rate_nearby_clusters,graham10:SNIa_rate,sharon10:hst_cluster_rates,dilday10:sdss_clusters}. 
Analysis
of these rates (Maoz et al. 2010b) reinforces the conclusion that, although the SN~Ia DTD may have a low-level tail at
delays of a few Gyr, the bulk of the events must occur within $\sim 1$~Gyr of star formation.

Finally, \cite{maoz10:SN_DTD_LOSS} have presented a method for recovering the DTD from a SN survey, where the SFHs of the
individual galaxies surveyed are taken into account. They applied the method to a subset of the Lick Observatory SN
Search (LOSS -- Filippenko in prep.; Leaman et al. in prep.; Li et
al. in prep.) that overlaps with the Sloan Digital Sky
Survey \citep[SDSS;][]{york00:SDSS_Technical}, for which SFH reconstruction is availale based on the SDSS spectroscopy
\citep{tojeiro09:SDSS_VESPA}. They found that a ``prompt'' ($<420$~Myr) SN~Ia component is required by these data at the
$>99\%$ confidence level. In addition, a delayed SN~Ia population, with delays of $>2.5$~Gyr, is detected at the $4\sigma$
level. A related approach has been used by \cite{brandt10:Ia_progenitors}, with similar 
conclusions.

In this paper, we seek to recover the SN DTD in yet another way, by analyzing the SNRs in the Magellanic Clouds
(MCs). The MC SNRs present several advantages.  First, they are all at the known distances of their two host
galaxies. Second, both MCs have been surveyed to large depths in the radio, with individual sources followed up at
multiple wavelengths and classified. In a companion paper (Badenes, 
Maoz, \& Draine 2010, hereafter Paper I), 
we have compiled a sample of MC SNRs, and argued that it is largely complete. Third,
the MCs are close enough to permit a region-by-region fitting of their resolved stellar populations with detailed
stellar evolution isochrones. The SFHs of the Clouds can therefore be reconstructed with better spatial and temporal
detail than in any other galaxies.  \citet{harris04:SMC_SFH} and \citet{harris09:LMC_SFH} have recently carried out such
a program of SFH reconstruction of the MCs. \citet{badenes09:SNRs_LMC} have compared the SFHs of the MC regions hosting
specific SNRs to the properties of the remnants, in order to deduce constraints on the nature of some of the explosions
and their delay times.

Here, we go a step further, and treat the MCs and their SNRs as an effective SN survey conducted in a sample of galaxy
subunits, where the detailed SFH of each subunit is known.  In \S2, below, we briefly review the reconstruction of MC
SFHs by \citet{harris04:SMC_SFH} and \citet{harris09:LMC_SFH}.  In \S3
we review our MC SNR sample from Paper I, and
the physical model that reproduces the observed size distribution of
this SNR sample. We show how this same model permits
estimating the relative visibility times of SNRs at different locations in the MCs, given the local ambient
densities. As in Paper I, we use three different tracers of MC density
to estimate the SNR visibility times as a function of location. 
In \S4, we review the method of \cite{maoz10:SN_DTD_LOSS} for recovering the most likely
SN DTD and its uncertainty, given the spatially resolved SFHs, the SNR visibility times, and the observed number of
SNRs. In \S5 we apply the method to the MCs with their sample of SNRs,
and derive the DTD.
In \S6 we use the visibility times to obtain also the current SN rate in
the MCs, and discuss the emerging picture.

 For the purpose of this paper, we assume that the
DTD is a universal function: it is the same in all galaxies,
independent of environment,
metallicity, and cosmic time --- a simplifying assumption that may be
invalid at some level. For example, a dependence of SN delay time 
on metallicity is 
expected in some models (e.g., Kobayashi et al. 2000). Similarly,
variations in the initial mass function (IMF) with
cosmic times or environment would also lead to a variable DTD,
but we will again ignore this possibility in the present context.

\section{The star-formation history maps of the Magellanic Clouds}
\label{sectionHZ}

The SFH maps that we use in the present work are presented in full
 detail in
\citet{harris04:SMC_SFH} and \citet{harris09:LMC_SFH}\footnote{The complete maps are available at
  \url{http://ngala.as.arizona.edu/dennis/mcsurvey.html}.}. The maps were elaborated using four-band (\textit{U, B, V},
and \textit{I}) photometry from the Magellanic Clouds Photometric Survey \citep{zaritsky04:MCPS}, which has a limiting
magnitude between 20 and 21 in \textit{V}, depending on the local degree of crowding in the images. In each Cloud, the
data were divided into regions or ``cells'' with enough stars to produce color-magnitude diagrams of sufficient quality,
which were then fed into the StarFISH code \citep{harris01:StarFISH}
 to derive the local SFH for each cell. For the Large Magellanic Cloud
 (LMC),
\citet{harris09:LMC_SFH} divided more than 20 million stars into spatial cells encompassing the central
8$^\circ\times8^\circ$ of the galaxy (see their figure 4). The
majority of the cells are $12'\times12'$ squares, while about 50 cells
in regions of lower stellar density have sizes of  $24'\times24'$. 
In total, there are 1376 cells for the LMC, for which the SFH is given in 13 temporal bins with
lookback times between 6.3 Myr and 15.8 Gyr. For the Small Magellanic
 Cloud (SMC), \citet{harris04:SMC_SFH} divided over 6 million stars into 351
$12'\times12'$ cells, leaving out two areas that are contaminated by Galactic globular clusters in the foreground (see
their figure 3). The temporal binning of the SFHs is slightly different for the SMC, with 18 resolution elements between
4.6 Myr and 9.7 Gyr. In the original maps, the SFHs are further subdivided into metallicity bins, but we ignore this
dimension, working with the co-added SFHs instead.

The stellar masses formed in every time bin in every cell, as
derived by \citet{harris04:SMC_SFH} and \citet{harris09:LMC_SFH},
assume a \citet{salpeter55:IMF} IMF. For ease of comparison of our
results to other SN rate work, we convert these masses to 
 a ``diet Salpeter'' IMF \citep{bell03:IMF}, by multiplying the
masses formed by  factor of 0.7, to account for the reduced number of
 low-mass stars in a realistic IMF, compared to the original 
\citet{salpeter55:IMF} IMF.

\section{The Sample of Supernova Remnants in the Magellanic Clouds,
 its Properties, and Physics}

In Paper I, we assembled a list of 
77 SNRs in the MCs, by cross-identifying literature compilations of
SNRs. We showed that
 there is an observed ``floor'' of 50~mJy in the radio flux of the 
SNRs, orders of magnitude higher than the flux limits
of the radio surveys in which almost all of these SNRs are
detected. From this, we concluded that 
the paucity of SNRs below $\sim 50$~mJy must be real, rather than being due to any
observational effect.  This suggested that any MC SNRs with radio
fluxes 
under the $\sim 50$~mJy floor but above the
detection limit must fade very quickly through the flux range from the floor down to the detection limit.

Analysing our sample in Paper~I, we showed that the size distribution
of the SNRs is close to linear in the cumulative, or uniform in the
differential, up to a cutoff at a radius of $r\sim 30$~pc. Such a
distribution of SNR sizes has been noted before both in the MCs and in 
other galaxies. It has been variously attributed to observational
selection effects or to the SNRs being in a ``free-expansion''
phase (i.e., with constant shock velocity), expected when
the mass swept up by the expanding shock is still 
small compared to the ejected mass.
We suggested an alternative physical model to explain the MC SNR size distribution.
 
Briefly,
we proposed that most of the MC SNRs are in the decelerating
Sedov-Taylor phase, where the swept-up mass is larger than the ejecta
mass, but the cooling time of the shocked gas is still longer than 
the age of the remnant, and
hence the evolution is approximately adiabatic. 
Once the cooling time becomes comparable to the age, the SNR quickly
loses energy radiatively, entering the radiation-loss-dominated snowplough phase, after which it slows down, breaks up,
and merges with the interstellar medium.
Quantitative estimates show that typical SNRs should be in their Sedov-Taylor phases for ages between a few and a few
tens of kyrs, and for sizes of order a few to a few tens of parsecs
\citep[e.g.][]{cioffi88:Radiative_SNRs,blondin98:Radiative_SNRs,truelove99:adiabatic-SNRs}.
During this  phase, the SNR sizes grow as
\begin{equation}
\label{sedov}
r\sim E_0^{1/5}\rho^{-1/5}t^{2/5}, 
\end{equation}
where $E_0$ is the kinetic energy of the explosion, $\rho$ is the ambient gas density, and $t$ is the time.  The shock
velocity therefore decreases as
\begin{equation}
\label{sedovv}
v=\frac{dr}{dt}\sim E_0^{1/5} \rho^{-1/5}t^{-3/5},
\end{equation}
or equivalently expressed in terms of $r$ rather than $t$,
\begin{equation}
\label{vsimrhor}
v\sim E_0^{1/2} \rho^{-1/2}r^{-3/2}.
\end{equation}

We proposed in Paper~I that the uniform size distribution arises as a result of the transition from the Sedov phase to the
radiative phase.  The radius of this transition depends on ambient density. When coupled with the distribution of
densities in the MCs, this leads to fewer and fewer sites at which large-radius Sedov-phase SNRs can exist.
The cooling time of the shocked gas depends on the density as
\begin{equation}
\label{eqtcool}
t_{\rm cool}\sim  \frac{kT}{\rho \Lambda(T)}
\end{equation}
where $\Lambda(T)$ is the cooling function at temperature $T$. 
The cooling function can be approximated with a
power law, $T^\epsilon$, in the temperature range of relevance for the shocked gas, around
$10^6~{\rm K}$. By relating the temperature to the shock velocity $v$, $kT\sim m_p v^2$ (where $m_p$ is the proton
mass), and equating $t_{\rm cool}$ to the age $t$ as expressed in Eq.~\ref{sedovv}, one obtains that the transition
radius, $r_{\max}$, scales as
\begin{equation}
\label{rmaxwithepsilon}
r_{\rm max}\sim  E_0^{(3-2\epsilon)/(11-6\epsilon)}\rho^{-(5-2\epsilon)/(11-6\epsilon)}.
\end{equation}
A remnant expanding beyond this radius, at the given ambient density, will enter the radiative phase and quickly fade
from view.  For a fairly large range of plausible cooling function dependences, e.g., indices $\epsilon$ of $-1/2$ to
$-3/2$, $r_{\rm max}$ depends on density as $\rho^{-3/7}$ to $\rho^{-2/5}$. Conversely, the maximum ambient density that
will permit a Sedov-phase SNR of radius $r$ is
\begin{equation}
\rho_{\rm max}\propto  r^{\delta},
\label{rhomaxvsr}
\end{equation}
where $\delta$ is likely in the range $-7/3$ to $-5/2$.  
We showed in Paper~I that  
a uniform size distribution, $dN/dr\approx {\rm const.}$, is obtained if 
the ambient gas density follows a 
power-law distribution of density, with index $\beta\approx -1$.

We then examined three different tracers of
gas density in the MCs:
the HI column density;
the star-formation rate (SFR) based on resolved stellar populations,
translated to a density via an inverse Schmidt law;
and the SFR based on the H$\alpha$ emission-line luminosity, again
combined with an inverse Schmidt law.
We showed that, indeed, these tracers suggest that the density distribution behaves as a power law of slope $\sim
-1$, as hypothesized, over at least an order of magnitude in density,
lending support to our picture of the SNRs being largely in their
Sedov phases.

If indeed the end of the visibility of an SNR in the MCs is determined solely by its transition to the radiative phase,
we can, in principle, determine at every point in the Clouds, the ``visibility time'' of SNRs, i.e., the time during
which a SNR would be visible, if it were there. This variable, (often called the ``control time'' in SN surveys) is an
essential input to any SN rate calculation. As in Eq.~\ref{rmaxwithepsilon} for the transition radius, we can derive
the dependence of the transition time, $t_{\rm max}$, on the explosion energy and the ambient density,
\begin{equation}
\label{tmaxwithepsilon}
t_{\rm max}\sim  E_0^{(2-2\epsilon)/(11-6\epsilon)}\rho^{-(7-2\epsilon)/(11-6\epsilon)}.
\end{equation}
Again, for a cooling-function power-law dependence on temperature with index of $\epsilon=-1/2$ to $-3/2$, the
dependence of the visibility time on the density is in a limited range, from $\rho^{-4/7}$ to $\rho^{-1/2}$.

To calculate an accurate SN rate, one needs not only the dependence of the visibility time on the variables, but also
the constant of proportionality. For example, a factor 2 increase in visibility time translates directly to a decrease
by the same factor of the SN rate.  Unfortunately, it is impossible to know, except crudely, what is the absolute
transition time at some point in the MC. The identification of the transition to the non-adiabatic phase with the point
in time when the age equals the cooling time is merely an order-of-magnitude device. The proportionality constant in
Eq.~\ref{tmaxwithepsilon} will depend, among other things, on the approximate power-law slope $\epsilon$ chosen to
replace the true cooling function, the normalization of the cooling function, which is a function of the metallicity,
the Mach number of the shock, and its adiabatic index. Reasonable variations of those parameters alone could already
change the proportionality constant by an order of magnitude.  A fiducial value of $10^{51}$ erg is often assumed for $E_0$, the
initial kinetic energy of a SN explosion. However, there are few SNe or SNRs that have sufficient data {\it and} are at
accurately known distances such that $E_0$ can be reliably determined.  The distribution $P(E_0)$ is also uncertain,
although energy budget arguments can be made to argue that extreme deviations from the fiducial value should be rare in
both CC and Type Ia events. Because the dependence of $t_{\rm max}$ on $E_0$ is mild, roughly $E_0^{1/4}$ or weaker,
$P(E_0)$ should not be a major source of variations in the mean visibility time, but the precise magnitude of these
variations is difficult to quantify.  We conclude that, based on the
above analytic treatment, the mean transition time to the radiative stage
and hence the mean visibility time of SNRs is known only to
order-of-magnitude accuracy. (see, however, \S\ref{compvistime}.)

It would be hoped that the relation in Eq.~\ref{tmaxwithepsilon} could be calibrated empirically, e.g., if we had an
independent age estimate for a SNR of measured size, known to be in its Sedov phase. However, there is currently no way
to obtain reliable age estimates for the relatively large and old SNRs considered here \citep[although some constraints
can be derived from SNR-pulsar associations, see][and references therein -- we revisit this issue in
\S\ref{mcapply}]{gaensler95:PSR_SNR_Connection_I}. The discovery of light echoes associated with SNRs J0509.0$-$6844
(N103B), J0509.5$-$6731 (B0509$-$67.5) and J0519.6$-$6902 (B0519$-$69.0) in the LMC \citep{rest05:LMC_light_echoes} has
opened interesting possibilities for a few objects, but they are all too small and young to serve as calibrators. To
proceed with our calculation of the SN rate in the MCs, we will therefore simply leave $t_{\rm max,0}$, the transition
time out of the Sedov phase for a SNR at a location having the mean ambient gas density of the Clouds, as an adjustable
parameter.  At different locations in the MCs, we will scale the visibility time according to the density, as indicated
by various tracers at those locations, according to
\begin{equation}
\label{tmaxscaled}
t_{\rm max}=t_{\rm max,0}~\rho^{-4/7~{\rm to}~ -1/2}.
\end{equation}
This will give us the {\it relative} visibility time at each location, which will be useful in \S\ref{mcapply} below,
when comparing SNR numbers and star fomation histories in indvidual MC cells. The proportionality constant, $t_{\rm
  max,0}$ will be obtained by requiring that the CC-SN yield (i.e., the number of CC-SNe per unit stellar mass formed)
corresponds to the theoretical expectation that most massive stars undergo CC-SN explosions. This expectation is
supported by an independent estimate of the DTD of CC-SNe \citep{maoz10:SN_DTD_LOSS}.

\section{SNR Visibility times from three estimates of the gas density distribution in the Magellanic Clouds}
\label{densityestimates}

In Paper~I, we studied three different tracers of gas density in the MCs: the HI column density;
the star-formation rate (SFR) based on resolved stellar populations; and the H$\alpha$ emission-line luminosity.
We used these tracers to test successfully
the hypothesis that the flat distribution of SNR sizes results from a transition out of the Sedov phase, which is,
in turn, determined by the ambient density, combined with a density distribution that is approximately a power law of
index $-1$. We now use the same three
tracers of the gas density to scale the SNR visibility time. 

\subsection{21~cm emission-line-based HI column density}

 The surface brightness of HI 21~cm line emission in the MCs from the maps of
\citet{kim03:LMC_HI_Parkes_ATCA} and
\citet{stanimirovic99MNRAS.302..417S}, analysed in Paper~I,  is optically
thin, and hence directly proportional to the HI column density. Since the LMC posesses a fairly
face-on \citep[inclination $i\sim 35^{\circ}$,][]{vandermarel01:LMC_inclination}, well-ordered HI disk, the column
density should, in turn, be roughly proportional to the volume density $\rho$. 
Assuming that the mean volume density $\rho$ in a cell is proportional to its mean column density $N_H$, we can scale
the visibility time for each cell containing a SNR according to Eq.~\ref{tmaxscaled}, with $t_{\rm max,0}$ assigned to
cells having the mean $N_H$ of the clouds, $1.5\times 10^{21}~{\rm cm}^{-2}$ per cell.

\subsection{Schmidt Law plus star formation rates from resolved stellar populations}

As a second way to estimate the local gas densities, in Paper~I 
 we used the recent SFRs in each cell of the \citet{harris04:SMC_SFH} and
\citet{harris09:LMC_SFH} maps. Here, we use  this tracer of gas density
also to obtain the
scaling of the visibility times of SNRs in the MCs.
We translate the 35-Myr-averaged SFR to a mass column, using an inverse \citet{schmidt59:SFR}
law. \citet{kennicutt98:SF_review} updated the Schmidt law, relating star formation rate surface
density, $\Sigma_{\rm SFR}$, to gas mass column $\Sigma_{\rm gas}$, as
\begin{equation}
\Sigma_{\rm SFR}=(2.5\pm0.7)\times 10^{-4}\left(\frac{\Sigma_{\rm
    gas}}{M_\odot~{\rm pc}^{-2}}\right)^{1.4\pm0.15}M_\odot ~{\rm yr}^{-1}
{\rm kpc}^{-2}.
\label{schmidtlaw}
\end{equation}
Recent measurements and discussions of the Schmidt-Kennicutt relation
  can be found, e.g., in \cite{bigiel08:SF_law_sub_kpc}, and references
  therein.
As emphasized by these studies, the Schmidt law has a threshold at
  some mass column, 
below which the star-formation rate
  falls steeply. \cite{kennicutt89:SFR_galdisks}, for example, finds a threshold at a mass column of $\Sigma_{\rm
  gas}=3-4~M_\odot {\rm pc}^{-2}$, but threshold values several times higher
  have also been reported.
 From Eq.~\ref{schmidtlaw} with
$\Sigma_{\rm gas}=3.5~M_\odot {\rm pc}^{-2}$, the threshold mass column corresponds to a SFR of $0.0014 ~M_\odot~ {\rm
  yr}^{-1} {\rm kpc}^{-2}$, or $4.3\times10^{-5}~ M_\odot~ {\rm yr}^{-1}$ per $12'\times 12'$ cell in the
\citet{harris09:LMC_SFH} maps of the LMC. For the $24'\times 24'$ cells in the LMC the threshold is of course 4 times
higher, and for the $12'\times 12'$ cells in the SMC, which is 20\% more distant, it is 1.44 times higher.
We therefore use the inverse Schmidt law with an exponent of $1/1.4=0.714$ to obtain the density in each cell, down to
this level of SFR. Below this SFR level, we assign to the cell a constant density corresponding to the threshold
level. The density, in turn, is translated to a visibility time according to Eq.~\ref{tmaxscaled}, with $t_{\rm max,0}$
assigned to cells having the mean SFR of the clouds, $3.3\times 10^{-4}~M_\odot ~{\rm yr}^{-1}$ per cell.

\subsection{Schmidt Law plus star formation rates from H$\alpha$ emission}

Rather than measuring SFR via the resolved stellar populations of the MCs, we can trace it by means of H$\alpha$
emission, which then gives us a third tracer of gas density. H$\alpha$ is principally powered by photoionisation from
O-type stars, whose numbers are proportional to the SFR. The Schmidt law then, again, connects SFR and gas mass
column. In Paper~I, we analyzed the continuum-subtracted H$\alpha$ emission maps of the LMC and SMC from the SHASSA survey of
\citet{gaustad01:SHASSA}, and showed that both the LMC and the SMC have distributions of H$\alpha$
surface brightness that follow power laws of index $-1$, over 2 orders
of magnitude in flux for the SMC, and over 3 order of magnitude for the
LMC. Through the Schmidt law, this implied a power-law with index $-1$ for the gas density distribution, over
2 orders of magnitude in density.
Here, we use the same H$\alpha$ maps and 
 translate the H$\alpha$ surface brightness to a visibility time via the SFR and the Schmidt law. We again use the
\citet{kennicutt89:SFR_galdisks} threshold of $4.3\times10^{-5}~ M_\odot~ {\rm yr}^{-1}$ per MC cell. By comparing the
mean H$\alpha$ surface brightness and the SFR in each cell, we find that this threshold corresponds to $\sim 100 {\rm dR}$
(deciRayleighs) in H$\alpha$ flux per typical MC cell (this holds for
all cells, both in the LMC and the SMC, since surface brightness is
independent of distance for these nearby galaxies).
Below this H$\alpha$ flux, we assign to a cell, in the absence of a density indicator, a constant density corresponding
to the threshold level.

\section{Reconstruction of the SN Delay Time Distribution -- Method}

With our SNR sample, and our estimate for the visibility time of SNRs at each location in the MC in hand, we now briefly
present a method to combine this information with the localized SFHs for each individual cell in the MC, in order to
recover the SN DTD.  A more detailed exposition of the method, including demonstrations of its performance on simulated
datasets, is presented in \cite{maoz10:SN_DTD_LOSS}.

The SN rate in a galaxy observed at cosmic time $t$ is given by the convolution
\begin{equation}
\label{convolution}
r(t)=\int_0^{t} S(t-\tau)\Psi (\tau) d\tau ,
\end{equation}
where $S(t)$ is the star-formation rate (stellar mass formed per unit time), and $\Psi (\tau)$ is the DTD (SNe per unit
time per stellar mass formed). As reviewed is \S1, previous attempts to recover the DTD have used rates $r(t)$ measured
in surveys of galaxies at different redshifts (i.e., different cosmic times), compared to cosmic star-formation
histories, whether in field surveys or galaxy cluster surveys.  An alternative approach has been to look at the SN rates
per unit stellar mass in galaxies of particular types (star-forming, quiescent, etc.), and to attempt to assign to each
type a ``formation age'' or some generic, simple, star-formation history. A shortcoming of all these approaches is that
they involve averaging over the galaxy population (i.e., all the SNe are assumed to come from the entire host population
considered), or over time (i.e., the detailed history of all galaxies
of a certain type is represented by a single ``age'' or
simplified history). As a result, all of these approaches involve loss of information, and potential systematic errors
(e.g., due to unrepresentative simplified histories).
 
An alternative way to recover the DTD is by inverting a linear, discretized version of Eq.~\ref{convolution}, where the
detailed history of every individual galaxy or galaxy subunit is taken into account. Suppose the star-formation
histories of the $i=1,2,...,N$ galaxies or galaxy subunits monitored as part of a SN survey are known (e.g., based on reconstruction of
their stellar populations), with a temporal resolution that permits binning the stellar mass formed in each galaxy into
$j=1,2,..,K$ discrete time bins, where increasing $j$ corresponds to increasing lookback time.  The time bins need not
necessarily be equal, and generally will not be, since the temporal resolution of the star-formation-history
recontruction degrades with increasing lookback time.  For the $i$th galaxy in the survey, the stellar mass formed in
the $j$th time bin is $m_{ij}$.  The mean of the DTD over the $j$th bin (corresponding to a delay range equal to the
lookback-time range of the $j$th bin in the star-formation history) is $\Psi_j$. Then the integral in
Eq.~\ref{convolution} can be approximated as
\begin{equation}
\label{discreteconvolution}
r_i\approx\sum_{j=1}^K m_{ij}\Psi_j  ,
\end{equation}
where $r_i$, the SN rate in a given galaxy, is measured at a particular cosmic time (e.g., corresponding to the redshift
of the particular SN survey). Given a survey of $N$ galaxies, each with an observed SN rate $r_i$ and a known binned
star-formation history $m_{ij}$, one could, in principle, invert this set of linear equations and recover the best-fit
parameters describing the binned DTD: ${\bf\Psi}=(\Psi_1, \Psi_2,...,\Psi_K$).

In practice, on human timescales SNe are rare events in a given galaxy, i.e., $r_i\ll 1 ~{\rm yr}^{-1}$. Supernova
surveys therefore monitor many galaxies, and record the number of SNe discovered in every galaxy. For a given model DTD,
$\Psi_1,..,\Psi_K$, the $i$th galaxy will have an expected number of SNe
\begin{equation}
\lambda_i=r_i t_i ,
\label{eqlambda}
\end{equation}
where $t_i$ is the effective visibility time during which a SN would have been
visible (given the actual on-target monitoring time, the distance to the galaxy, the flux limits of the survey, and the
detection efficiency). Since $\lambda_i \ll 1$, the number of SNe observed in the $i$th galaxy, $n_i$, obeys a Poisson
probability distribution with expectation value $\lambda_i$,
\begin{equation}
P(n_i|\lambda_i)=e^{-\lambda_i}\lambda_i^{n_i}/n_i ! ,
\end{equation}
where $n_i$ is zero for most of the galaxies, 1 for some of the galaxies, and more than 1 for very few galaxies.

Considering a set of model DTDs, 
 the likelihood of a particular DTD, given the set of measurements
 $n_1,...,n_N$, is
\begin{equation}
L=\prod_{i=1}^N P(n_i|\lambda_i).
\end{equation}
More conveniently, the log of the likelihood is
\begin{equation}
\ln L =\sum_{i=1}^N \ln
P(n_i|\lambda_i)
=-\sum_{i=1}^N{\lambda_i}
+\sum_{i=1}^N\ln (\lambda_i^{n_i}/n_i !), 
\end{equation}
where obviously only galaxies hosting SNe contribute to the second term.
The best-fitting model can be found by scanning the parameter space
of the vector ${\bf \Psi}$ for the value that maximizes the log-likelihood.
This procedure naturally allows restricting the DTD to have
only positive values, as physically required 
(a negative SN rate is meaningless).

The covariance matrix $C_{jk}$ of the errors in the best-fit parameters can be found by calculating the curvature
matrix,
\begin{equation}
\alpha_{jk}=\frac{1}{2}\frac{\partial^2\ln L}{\partial\Psi_j \partial\Psi_k}
=\sum_{i=1}^N\frac{\partial[\ln P(n_i|\lambda_i)]}{\partial\Psi_j}
\frac{\partial[\ln P(n_i|\lambda_i)]}{\partial\Psi_k},
\end{equation}
and inverting it,
\begin{equation}
[C]=[\alpha]^{-1} .
\end{equation}
Because the values of the DTD are constrained to be positive, if the maximum-likelihood of a DTD component $\Psi_j$ is
close to zero, the square-root of its variance, ${\sqrt{C_{jj}}}$, will not represent well its $1\sigma$ uncertainty
range. An alternative, more reliable, procedure is to perform a Monte-Carlo simulation in which many mock surveys are
produced, each having the same galaxies, star-formation histories, and visibility times as the real survey, and having
expectation values $\lambda_i$ based on the best-fit DTD, but with the number of SNe in every galaxy, $n_i$, drawn from
a Poisson distribution according to $\lambda_i$. The maximum-likelihood DTD, ${\bf\Psi}$, is found for every
realization. From the distribution of the values of every component, $\Psi_j$, over all the realizations, one can
estimate the range encompassing, say, $^+_-34\%$ of the cases.
Another advantage of the Monte-Carlo simulations is that they allow
gauging the effects of additional sources of error, such as
uncertainties in the SFHs.

The above approach for recovering the DTD has several advantages over previous methods. First, all the known information
in the survey is included in the analysis in a statistically rigourous way, including the fact that many (usually most)
of the galaxies, or galaxy subunits, did not host any SNe. Furthermore, the calculation is easily generalized to cases
where the galaxies are not all at the same distances (e.g., combinations of surveys done at different redshifts); one
simply needs to use the appropriate star-formation history bins for every galaxy. In fact, assuming the DTD is a
universal function (i.e., it is independent of environment,
metallicity, cosmic time, etc., see \S1) it is straightforward to combine,
in a single analysis, the data from completely disparate SN surveys, e.g., classical SN surveys, where a large volume or
a lare sample of galaxies is monitored, with unconventional SN ``surveys'' such as the one presented here, in which the
SN rate is measured based on SN remnants in small subunits of a single
galaxy. We re-emphasize that, in the preceding discussion, everything relating to
``galaxies'' is equally applicable to individual subunits of a galaxy,
for which a SFH is available.

As described in more detail in \cite{maoz10:SN_DTD_LOSS},
the number and resolution of the time bins used in the analysis will naturally depend on the quality of the data; the
larger the number of observed SNe, $N_{\rm tot}$, the higher the time resolution that can be recovered with reasonable
accuracy.  For a survey with a fixed $N_{\rm tot}$, there will be a tradeoff in the analysis between DTD accuracy and
resolution.

\section{The DTD in the Magellanic Clouds}
\label{mcapply}
\subsection{General considerations}
We now apply the \cite{maoz10:SN_DTD_LOSS} DTD recovery method, 
described in the
previous section, to the 77 SNRs in the Clouds, treating the sample as a SN survey with visibility times
given by Eq.~\ref{tmaxscaled}, conducted over the 1836 galactic subunits or ``cells'' defined by
\citet{harris04:SMC_SFH} and \citet{harris09:LMC_SFH}, each with its reconstructed SFH, based on the resolved stellar
populations. We recall that changing the mean visibility time, $t_{\rm max, 0}$, which still needs to be adjusted,
causes the amplitude of the DTD to scale identically across all time
bins,
as  $t_{\rm max, 0}^{-1}$.

One point of concern in the derivation of the DTD could be that, because of the random stellar velocities in each
galaxy, the stars currently in a given cell which contributed to the SFH are not the same stars that produced the SNe
observed to have exploded in that cell, many years after the formation of their progenitors. While this is true, it is
inconsequential to a correct determination of the global DTD, because the same spatial diffusion affects both the
progenitor population and those stars among it that eventually explode.  To see this, consider, as an example, a grid of
$3\times 3$ adjacent cells in the LMC. Suppose, as a toy example, that 500~Myr ago there was a short burst of star formation
in the central cell, forming a stellar mass $M$, and no activity in the other cells. Suppose, further, that the SN DTD
is such that the stellar population formed in the burst leads, 500~Myr later, to nine SNe over the past 20~kyr, which
are therefore detected as SNRs in the galaxy, i.e., a ratio of $9/M$ SNe per unit stellar mass formed 500~Myr ago.
Finally, suppose that the stellar diffusion timescale in the galaxy is such that, over the 500~Myr, the progenitors of
the 9 SNe, before exploding, have drifted out of the central cell in which they were formed, and there is now, on
average, one SN in each cell. However, the entire stellar population of the burst will have diffused in the same way,
and therefore each cell will have 1/9 of the 500~Myr old population that was originally in the central cell. When we
compare SN numbers to the 500~Myr-old stellar mass present today in each cell, we will see 1 SN per $M/9$ of stellar
mass formed, and therefore we will still deduce the correct ratio of $9/M $ SNe per unit stellar mass formed. This
argument holds no matter what are the lookback times, 
the diffusion timescales, or the complex SFH in each cell.

A peculiarity of our SN survey is that, for most of the remnants, we have little or no information on the type of the SN
that exploded. Some young SNRs can be classified using different
methods: SNR 1987A was obviously a CC-SN, SNR J0509.5$-$6731
(B0509$-$67.5) was a SN~Ia both from the spectroscopy of its light echo \citep{rest08:0509} and the X-ray emission from
the SN ejecta \citep{badenes08:0509}, and several other objects have classifications that range from the very secure
(those harbouring compact objects are almost certainly CC, see below) to the reasonable (young objects with
ejecta-dominated X-ray emission). However, these methods usually cannot be applied to the old SNRs in the Sedov phase
that form the bulk of our objects \citep[for a more detailed discussion on SNR typing, see \S~3
in][]{badenes09:SNRs_LMC}.  The SNR sample therefore constitutes a mix of different SN types. However, this is easily
addressed within our formalism.  As already noted above (see \ref{lifetimes}, CC SNe will generally explode within $t_{\rm CC}\sim 30-40$~Myr
of star formation, and SNe~Ia will explode after this time.  We will therefore choose our shortest time bin as $0<\tau
<35$~Myr, and interpret the DTD amplitude in this bin as the CC SN signal, while the signal in later bins is due to the
SNe~Ia.

A systematic error arises through the uncertainty in the visibility time, due to the the cooling function dependence on
temperature (Eqns.~\ref{tmaxwithepsilon}-\ref{tmaxscaled}), and due to the three different tracers of density that we
consider -- HI column density, resolved SFR, and H$\alpha$ luminosity (\S\ref{densityestimates}). We
probe the effect of this systematic uncertainty, by recovering the best-fit DTD for all combinations of the density
tracers with the two extreme density exponents of Eq.~\ref{tmaxscaled}: $-4/7$ and $-1/2$.  We find that the choice of
exponent affects the derived DTD amplitudes only slightly, by $< 3\%$ for the HI and H$\alpha$ tracers, and by 10\% for
the resolved star formation rate tracer.  We will henceforth cite results only for an exponent of $-1/2$.

\begin{figure*}
  \begin{center}
  \includegraphics[width=0.29\textwidth, angle=90]{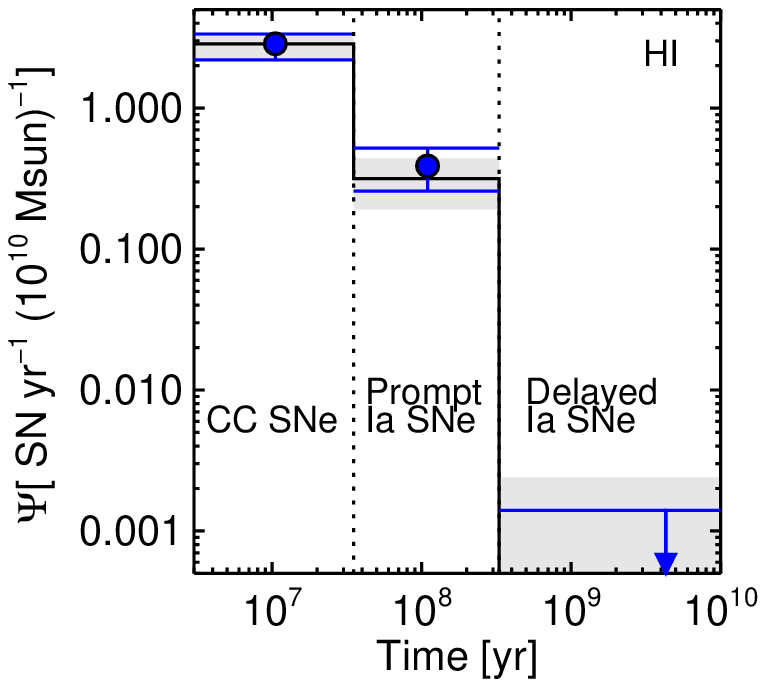}
  \includegraphics[width=0.29\textwidth, angle=90]{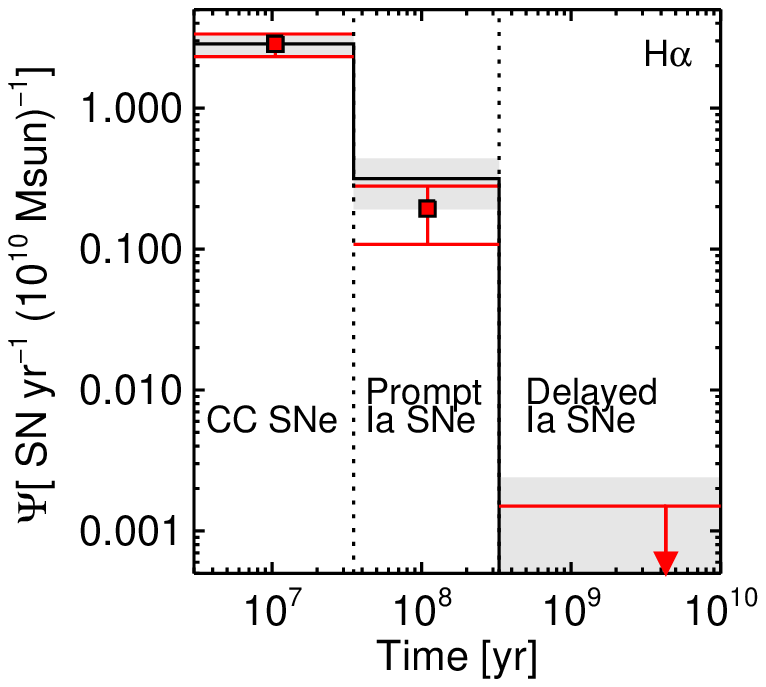}
  \includegraphics[width=0.29\textwidth, angle=90]{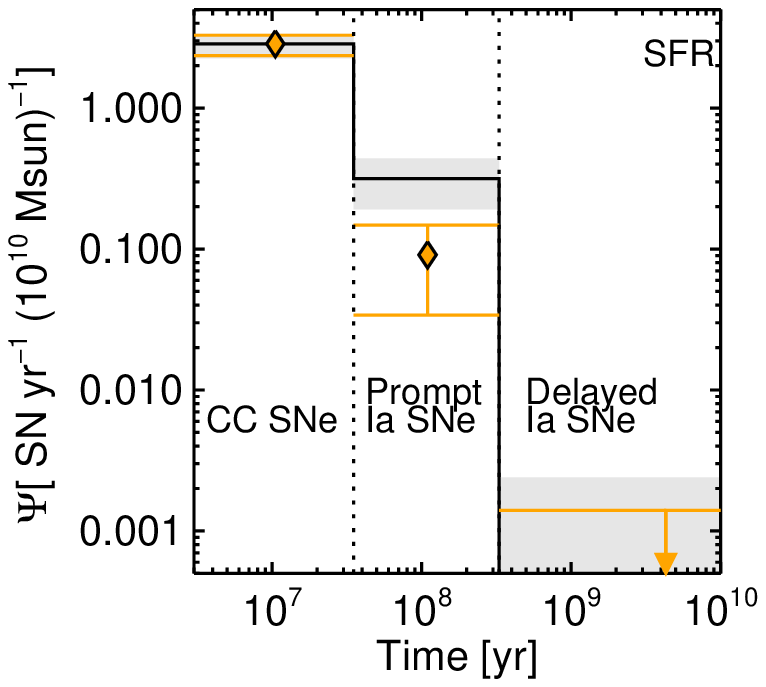}

  \includegraphics[width=0.29\textwidth, angle=90]{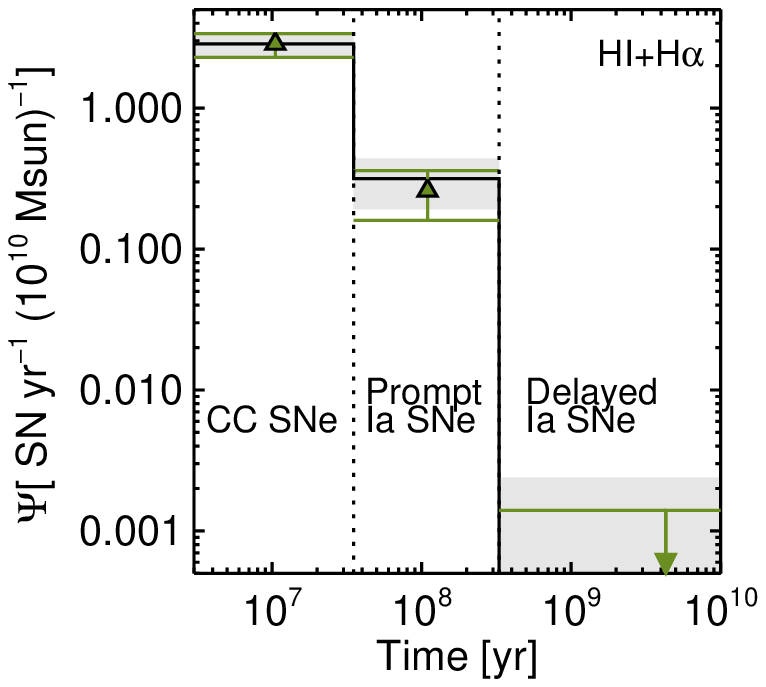}
  \includegraphics[width=0.29\textwidth, angle=90]{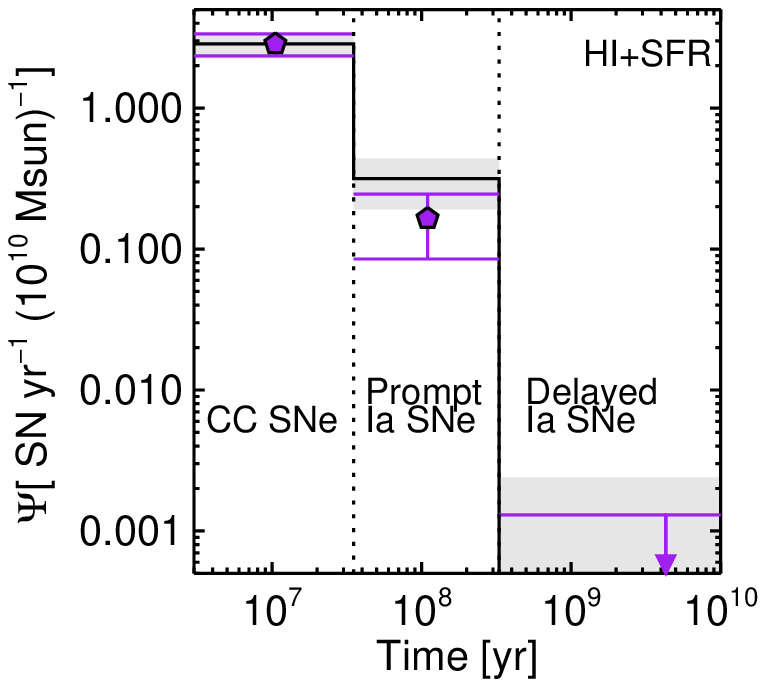}
  \caption{Best-fit delay-time distributions for the Magellanic Cloud SNR sample. Each panel in the top row displays the
    results obtained with a different density tracer: HI (left), H$\alpha$ (centre), and SFR (right). The panels in the
    bottom row show the DTDs obtained with ``hybrid'' density tracers, using HI at low densities and either H$\alpha$
    (left) or SFR (right) at high densities. Error bars show 68\% confidence intervals, from Monte-Carlo simulations
    that account both for Poisson statistics and for the uncertainties in the SFR of each cell. The best fits in the
    most-delayed bins are always zero, and the horizontal line gives the 95\% confidence upper limit on the rate in this
    time bin.  To facilitate comparison, the DTD obtained without scaling with density of the visibility time of SNRs is
    plotted in all panels as black lines, with shaded regions for the error bars. All DTDs have been scaled so as to
    give the same, theoretically expected, CC-SN yield in the first bin.}
  \label{lmcddt}
  \end{center}
\end{figure*}

Figure~\ref{lmcddt}  and Table~2 show the Magellanic Cloud DTD we recover from the SNR sample, binning the DTD into the following
three intervals: $\tau<35$~Myr (CC SNe), $35~{\rm Myr}<\tau<330$~Myr (``prompt'' SNe~Ia) and $330~{\rm Myr}<\tau<14$~Gyr
(``delayed'' SNe~Ia).
The small number of SNe in our ``survey'' precludes the possibility of finer binning and a detailed recovery of the DTD
shape. Nevertheless, we will see that some interesting results emerge even with this coarse binning scheme.

\subsection{Core-collapse SNe}
For each of the density tracers, we renormalise the visibility time at the mean density, $t_{\rm max,0}$, such that the
rate in the first DTD bin, which traces CC-SNe, is $\Psi_1=2.86 ~{\rm SNe~yr}^{-1}(10^{10}M_\odot)^{-1}$. Multiplied by
the width of the bin, 35~Myr, this then gives a time-integrated CC-SN yield (i.e., the number of CC-SNe per unit stellar
mass formed) of $N_{\rm CC}/M=0.01 ~{\rm M_\odot}^{-1}$.  This is the value expected if all stars above $8 M_\odot$
explode as CC SNe,
\begin{equation}
\label{frac2}
\frac{N_{\rm CC}}{M}
=\frac{\int_{8}^{100}(dN/dm) dm}{0.7\int_{0.1}^{100}m (dN/dm) dm}=0.01 ~{\rm M_\odot}^{-1},
 \end{equation} 
 for a ``diet Salpeter'' IMF \citep{bell03:IMF}, where $dN/dm\propto
 m^{-2.35}$, $m$ is stellar mass in units of ${M_\odot}$, and the
 factor 0.7 in the denominator accounts for the reduced number of
 low-mass stars in a realistic IMF, compared to the original
 \citet{salpeter55:IMF} IMF (see \S\ref{sectionHZ}).

  As listed in Table~2 for the three density tracers, the renormalized visibility time at
 a location in the MCs having the mean density is: 22.5 kyr (resolved SFR); 13.9 kyr (H$\alpha$); and 13.3 kyr
 (HI). Depending on the local value of the density, individual SNRs in the MCs could obviously be much older -- at a
 density 10 times lower than the mean, for instance, the SNR lifetime would be a factor 3 to 4 longer, depending on the
 value of the exponent in Eq.~\ref{tmaxscaled}. Our SNR lifetimes are therefore in agreement with the maximum ages of
 SNR shells constrained by SNR-pulsar associations \citep[$\sim60$
 kyr,][]{frail94:SNR_lifetime}.

 Our resulting DTD will be incorrectly normalized if, in reality, the mass border between CC-SNe and SNe~Ia is not at
 $8M_\odot$, or if a large fraction of massive stars end their lives without a SN explosion that leaves a remnant
 \citep[e.g., through direct collapse to a black hole;][]{kochanek08:survey_nothing}.  However, a direct derivation by
 \cite{maoz10:SN_DTD_LOSS} of the CC-SN DTD from the LOSS-SDSS sample supports the conclusion that the CC-SN yield is
 at the level assumed here.

The errors we cite in Fig.~\ref{lmcddt} and
 Table~2 give the 68\% probability range of each component, based on Monte-Carlo
 simulations, as described above. In these simulations, 
to account also for the uncertainties given by \cite{harris09:LMC_SFH} and \cite{harris04:SMC_SFH} for their SFHs, we have done the following. 
For each MC cell, we add in quadrature the $1\sigma$ errors in SFH among the fine time bins that constitute a single
 co-added time bin in our analysis. Since there is covariance among the SFH errors in adjoining bins, this addition
 constitutes a conservative overestimate of the true errors. Then, in every mock survey in the simulation, we draw, for
 each cell in every time bin, a SFR from an asymmetric Gaussian distribution, centered on the \cite{harris09:LMC_SFH} and \cite{harris04:SMC_SFH} best-fit value, with positive and negative standard deviations according to the bin-added negative and positive
 errors above. After calculation of the expectation value of the rate for the cell, according to Eq.~\ref{eqlambda}, the
 ``observed'' number of SNRs in the cell is drawn from the appropriate Poisson distribution. An exception to this
 procedure is in the SMC, where we have avoided using the $1\sigma$ positive errors of \cite{harris04:SMC_SFH}, as
 they are two orders of magnitude larger than those given by \cite{harris09:LMC_SFH} for the LMC, despite the
 similarity between the two galaxies, their data, and their analysis.  For the SMC, we therefore use the negative errors
 in the SFH to represent the positive errors as well. Finally, in the
 DTD reconstruction of each mock survey, the Harris \& Zaritsky (2004,
 2009) best-fit value is assumed for the SFH (i.e., a value generally
 different than what was used to generate the mock survey).

 We note that, although we have renormalized $t_{\rm max,0}$, so as to always obtain $\Psi_1=2.86~ {\rm
   SNe~yr}^{-1}(10^{10}M_\odot)^{-1}$, $\Psi_1$ is still a parameter in the DTD reconstruction, and its derived error
 indicates the confidence level at which a CC-SN component is detected in the DTD. For the different density
 tracers, the $\Psi_1$ component of the DTD is detected in our reconstruction at the $5-7\sigma$ level. In other words,
 there is no doubt that some of the SNRs in the MCs are significantly
   associated with stellar populations that are young enough to
 produce CC-SNe.

 \begin{table*}
   \centering
   \begin{minipage}{165mm}
     \caption{Mean SNR Visibility times, SN rates, and reconstructed DTDs from the MC SNR sample.}
     \begin{tabular}{@{}lccccccc@{}}
       \hline
       Density  &  Mean SNR & \multicolumn{2}{c}{SN Rate}& \multicolumn{4}{c}{$\Psi_{i}$ [$\rm{SNe~yr^{-1}}\,(10^{10}\,\rm{M_{\odot}})^{-1}$]} \\
       Tracer & Visibility & & &
       $\Psi_{1}$\footnote{The rate in this bin was
       renormalized, but the errors reflect its significance, see text
       for details.} & \multicolumn{2}{c}{$\Psi_{2}$} & $\Psi_{3}$ \\
       & [kyr] & [$10^{-3}$ SN yr$^{-1}$] & [SNuM] &
       0-35 Myr  & 35-330 Myr & (95\%~low lim.)& 0.33-14
       Gyr \\
       \hline
       HI & 13.3 & $4.1\pm0.9$ & $3.0\pm0.7$ & $2.86^{+0.66}_{-0.50}$
       & $0.389\pm0.131$ & ($>0.171$)& $< 0.0014$ \\
       SFR (Schmidt law) & 22.5 & $2.4\pm0.4$ & $1.7\pm0.3$ & $2.86^{+0.50}_{-0.43}$ & $0.091\pm0.057$ & ($>0.016$)&$<0.0014$ \\
       H$\alpha$ (Schmidt law) & 13.9  & $3.3\pm0.6$ & $2.3\pm0.4$ & $2.86^{+0.54}_{-0.50}$ & $0.194\pm0.086$ & ($>0.060$)&$<0.0016$ \\
       No Scaling & 13.4 & $5.8\pm0.1$ & $3.5\pm0.7$ &
       $2.86^{+0.61}_{-0.50}$ & $0.316\pm0.125$ & ($>0.143$)&$<0.0024$ \\
       HI+H$\alpha$& 12.6 & $3.8\pm0.8$ & $2.6\pm0.5$ &
       $2.86^{+0.57}_{-0.52}$ & $0.260\pm0.100$ & ($>0.104$)&$< 0.0014$ \\
       HI+SFR& 15.9 & $3.1\pm0.6$ & $2.1\pm0.4$ & $2.86^{+0.51}_{-0.51}$ & $0.165\pm0.080$ &($>0.058$)& $< 0.0013$ \\
       \hline 
     \end{tabular}
SN rates in units of SN~yr$^{-1}$ are for the LMC and SMC together. SN
     rates per unit stellar mass are in SNuM [SNe~$(100~{\rm
     yr})^{-1}(10^{10}M_\odot)^{-1}]$, 
     after converting formed mass to present mass, by assuming 30\%
     mass loss.    
     Quoted numbers are: best-fit values and 68\% confidence intervals for the rates, 
$\Psi_1$ and $\Psi_2$; 
95\% confidence lower limits for
     $\Psi_2$;
and 95\% confidence upper limits for
     $\Psi_3$.
   \end{minipage}
 \end{table*}

\subsection{Prompt SNe Ia}
 The strong DTD signal recovered in the CC bin effectively ``uses up'' much of the SN data in the survey, leaving a
 weaker signal in the SN~Ia bins. Nevertheless, there is a signal in the ``prompt'' SN~Ia bin, $\Psi_2$, at
 35~Myr$<\tau<330$~Myr, at the $1.7\sigma$ level for the resolved star formation tracer, $2.3\sigma$ for the
 H$\alpha$ tracer, and $3\sigma$ when using HI column as the density
 tracer. The rate $\Psi_2$ itself is also
 correspondingly higher or lower when using each of these tracers, as
 shown in Table~2. Here, by ``$X\sigma$'' we mean the number of
 $-34\%$ uncertainty intervals above zero. Because of the non-Gaussian
 nature of the errors, the significance of the detection is higher
 that it would be for the same number of standard deviations in a
 Gaussian. Specifically, we find from Monte-Carlo simulations, in which 
we input a value of $\Psi_2=0$, that recovered values as high as
the best-fit values are
 obtained in fewer that 1\% of mock surveys, using any of the tracers.
In other words, a nonzero prompt SN Ia component is detected at $>99\%$
 confidence. Using additional simulations, we have determined, for
 each tracer, the input $\Psi_2$ rate that gives recovered $\Psi_2$
 values  greater of equal to the best-fit values in 5\% of the
 realizations. This input  $\Psi_2$ rate is the 95\% confidence lower
 limit on $\Psi_2$, which is also listed in Table~2.
 Despite the clear detection of a prompt SN~Ia
 component, there is factor 4 systematic uncertainty in the best-fit amplitude of this component.  This stems from the
 systematic uncertainty in estimating the local density, which in turn leads to an uncertainty in the scaling of the
 visibility time. 

 The $\Psi_2$ rate, multiplied by the length of the second time bin (295 Myr), indicates a time-integrated production of
 prompt SNe~Ia of: $N_{\rm Ia}/M=0.0027\pm 0.0014 ~{\rm M_\odot}^{-1}$ (resolved SFR); $N_{\rm Ia}/M=0.0057\pm 0.0026
 ~{\rm M_\odot}^{-1}$ (H$\alpha$); or $N_{\rm Ia}/M=0.0114\pm 0.0039 ~{\rm M_\odot}^{-1}$ (HI).  We note that $N_{\rm
   Ia}/M$, above, is essentially the ``$B$ parameter'' discussed by \citet{scannapieco05:A+B_models},
 \citet{mannucci06:two_progenitor_populations_SNIa}, and \citet{sullivan06:SNIa_rate_host_SFR}, i.e., the ratio of SN~Ia
 rate to SFR in strongly star-forming galaxies. As summarised in
 \citet{maoz08:fraction_intermediate_stars_Ia_progenitors}, the values of $B$ generally found by SN surveys are in the
 range $B=0.001$ to $0.003 ~{\rm M_\odot}^{-1}$, an exception being \citet{sullivan06:SNIa_rate_host_SFR}, who obtain a
 value several times lower, $(3.9\pm 0.7)\times 10^{-4}{\rm
 M_\odot}^{-1}$. 
The values we find for $N_{\rm Ia}/M$ are broadly consistent with those estimates,
 but tend to be on the high side, particularly when we use HI column as a density tracer in the MCs.

 The time-integrated ratio of CC-SNe to SNe~Ia (or equivalently, the ratio of rates of CC and type-Ia SNe in a
 population that has had a constant star-formation rate for a long time, such that both rates have reached a steady
 state) is in the range of $9:1$ to $1:1$, when we consider both the statistical and the systematic
 uncertainties.

\subsection{Delayed SNe Ia}
 The DTD amplitude in the third bin has a best-fit value of zero in all of our reconstructions, but an uncertainty range
 that includes the typical SN~Ia rates that have been measured in old, quiescent populations.  The uncertainty is large
 because few SNe are expected in this bin in our small sample, given the dominance of young stellar populations in the
 MCs.  Specifically, from our Monte-Carlo simulations, the $95\%$
 confidence
 upper limit on the delayed SN~Ia rate in the DTD is
 $\Psi_3<0.0013~ {\rm to}~ 0.0016$~SNe~yr$^{-1} (10^{10} M_\odot)^{-1}$
 (depending on the density tracer).  In other words, such an input value in the simulations for this
 component of the DTD results in a best-fit reconstructed value of zero in 5\% of cases. For comparison, the typical
 values found for the ``$A$ parameter'', the SN~Ia rate per unit mass in an old population that has no ongoing star
 formation, have been in the range $A\sim (2-10)\times 10^{-4}$ SNe yr$^{-1} (10^{10}M_\odot)^{-1}$ \citep[see
 compilation in][]{maoz08:fraction_intermediate_stars_Ia_progenitors}.  However, $\Psi_3$ is the rate per unit mass {\it
   formed}. The published rates in old populations are a per unit of {\it existing} stellar mass, i.e., in stars and
 stellar remnants. In the course of $\sim 10$~Gyr, a stellar
 population (with a ``diet Salpeter'' IMF that we are assuming
 throughout) will return $\approx 50\%$ of its mass to the interstellar
 medium via SN explosions and mass loss during stellar evolution \citep{bruzual93:isochrones}. Thus, for comparison to
 $\Psi_3$, $A$ needs to be multiplied by $0.5$, giving $\sim (1-5)\times 10^{-4}$ SNe yr$^{-1}$ per
 $(10^{10}M_\odot)^{-1}$ formed. This range fits comfortably, by at least a factor of 2, below the upper limit we have
 derived here based on the SNRs in the MCs. An application of our DTD recovery method to the LOSS SN survey, which does
 have a significant old galaxy population, by \cite{maoz10:SN_DTD_LOSS}, gives a highly significant measurement of a delayed
 SN~Ia component, $\Psi_3=2.6^{+0.8}_{-0.6}\times 10^{-4}$ SNe yr$^{-1} (10^{10}M_\odot)^{-1}$.  This value is
 consistent both with previous determinations of the $A$ parameter and with the upper limit on $\Psi_3$ found here.
 
\subsection{Relative fraction of prompt and delayed SNe Ia}
 Bins 2 and 3 of the DTD thus, respectively, require the existence of prompt SNe~Ia, and allow (but do not require) the
 existence also of a delayed component. We note that, if we choose to ignore the density distribution, and assume a
 constant visibility time throughout the MCs, we obtain results that are intermediate to those using H$\alpha$ and HI,
 in terms of visibility time and prompt-SN-Ia significance and amplitude,
 and hence also in terms of SN~Ia yield and CC-SN to SN~Ia time-integrated
 ratio.

We can also use the $\Psi_2$ measurements and the 
upper limits on $\Psi_3$ to gauge the relative fraction of the
prompt and delayed SNe among the SN~Ia population.
Using the resolved-SFR-based values of $\Psi_2$ and $\Psi_3$ in
Table~2, which have the lowest ratio, we can obtain a lower limit 
on the time-integrated SN~Ia yields in the two components,
\begin{equation}
\frac{\Psi_2 \Delta t_2}{\Psi_3 \Delta t_3}\gtrsim 1,
\end{equation}
wher $\Delta t_2$ and $\Delta t_3$ are the sizes of time-bins 2 and 3.
Using the other tracers gives lower limits on this ratio that are larger. In any
case, a significant fraction  of SNe~Ia, possibly a majority, are prompt.

\subsection{Hybrid density tracers}
 Among the three density tracers that we use, the HI column density is
 physically closest to the actual volume density we are interested in.  Furthermore, it can probe and represent the
 lower-density regions of the MCs, where the two other tracers, which rely on an inverse Schmidt law, break down because
 of the mass column threshold in the SFR. In those low-density regions, we have knowingly assigned a constant,
 artificially high, density which systematically shortens the visibility time. Finally, the Schmidt law has a large
 intrinsic scatter, inducing noise in the assignment of visibility times. If we focus on the HI-tracer results, we find
 a highly significant prompt-SN-Ia component, with an amplitude somewhat higher than found by field SN surveys, and a
 time-integrated ratio of CC-SNe to SNe~Ia of approximately 1:1. Interestingly, by looking at this ratio for the youngest
 remnants in the LMC, which are the ones that can be most securely classified, one would have guessed as much -- four of
 the eight smallest SNRs in the LMC were definitely or likely type Ia's \citep{badenes09:SNRs_LMC}. 

 On the other hand, HI as a density tracer also has its shortcomings.
 For example, the cell in the centre of the 30 Doradus region of the LMC has
 a high SFR, as evidenced by its resolved mean SFR over the last 35~Myr, which is 5 times higher than the mean of the
 MCs, and by its H$\alpha$ luminosity, which is 30 times higher than the mean. The HI column density, however, is only
 twice as high as the mean for the MCs. \citet{harris09:LMC_SFH} have already noted the far-from-perfect spatial
 correlation amongst different tracers of star formation. In the case of 30 Doradus, the HI column is likely
 underestimating the true total gas density, because a fraction of the neutral hydrogen has been photoionised by the
 massive stars in this region. An underestimate of gas density in such regions will lead to an overestimate of the
 visibility time. This, in turn, will lower $\Psi_1$, the DTD rate of CC-SNe that are associated with such regions.
 Renormalising the entire DTD such that $\Psi_1=2.86$ will then lift also the prompt-SN-Ia component, $\Psi_2$, causing its
 rate to be overestimated. At some level, this is likely occurring when we use HI column as a density tracer. Furthermore,
 we have discussed in Paper~I the evidence that HI may be misrepresenting the density in the SMC because of projection
 effects. Thus, this tracer is also not foolproof.
 
To deal with these problems, we have combined two different
tracers, using each in the regime where it is likely to be most
reliable. We have rederived the visibility time in each cell, using HI column
as the density tracer in cells where the column density is lower
than the mean (about 2/3 of the cells), and using SFR and the inverse
Schmidt law, as before, in the rest of the cells, where HI is above
the mean. The SFR is traced either with the resolved stellar
populations, or using H$\alpha$.
 
The best-fit DTD recovered with these visibility times (see Table~2), not
surprisingly, is intermediate to the ones found with HI and H$\alpha$
or the resolved SFR
when used
alone as tracers. The prompt SN~Ia component is detected at a level of
$2.6\sigma$ above zero (using HI + H$\alpha$) or 
$2.1\sigma$ above zero (using HI + resolved SFR). As before,
from the Monte Carlo simulations, the significance of these detections is $>99\%$.
The results do not depend strongly on the adopted border for the use
of the two tracers. For example, setting the border between use of HI and
resolved SFR at 1/2 of the mean HI column or at twice the mean HI
column (instead of at the mean), changes the best-fit value of
$\Psi_2$ by $-27\%$ and $+20\%$, respectively.
We consider these hybrid-tracer-based DTDs to be our most reliable results.

\subsection{Sensitivity to temporal binning and leaks}
 A concern with all of our results is that they may be sensitive to the temporal border between CC-SNe and SNe~Ia, which
 we have set at 35~Myr.  For example, if some CC-SNe explode, say, between 35~Myr and 45~Myr, some of the CC-SN signal
 will be attributed incorrectly to SNe~Ia. As a result, we will
overestimate the number of prompt SNe~Ia. To test this possibility, we have performed the following
 experiment. We have selected 11 LMC SNRs from Paper~I that are almost certainly remnants of CC explosions, based
 on one or more of the following criteria: detection of pulsars within them, X-ray analysis of the ejecta showing clear
 CC-SN products, or (in the case of SN1987A) direct historical evidence. With this small sample of known CC-SNe, which
 we list in Table~3, we have re-derived the DTD. Naturally, the amplitude of the shortest, CC-SN, bin of the DTD will be
 understimated, because we have selected only a fraction of the CC remnants in the LMC. Nevertheless, we can test if any
 signal ``leaks'' into the later bins, which we have been associating with SNe~Ia, either because of the time border
 issue discussed above, or because of some other systematics of the \citet{harris09:LMC_SFH} SFH
 reconstruction. We find no evidence of ``leakage'' outside the
 $<35$~Myr bin. For example, using the HI+H$\alpha$ hybrid density
 tracer, the best-fit results are, in the
 first (earliest) time bin of the DTD (with no renormalisation of
 this rate), 
$\Psi_1=0.61\pm 0.26$~SNe~yr$^{-1} (10^{10} M_\odot)^{-1}$, and zero in the second and
 third bins. This is reassuring, but because of the small size of this subsample, 
the 95\% confidence upper limit on the rates
in the second and third bins are not low enough to rule out, using
 this test, a leak of the CC signal, having the amplitude of the second and
 third bins in the full sample. 

 As another test, we have rederived the full-sample DTDs with the various tracers, but enlarging the first time bin from
 0-35~Myr to 0-80~Myr (this step size is dictated by the time bins of the Harris and Zaritsky SFH reconstructions). Not
 surprisingly, the best-fit amplitude of $\Psi_2$ decreases by factors of 2-3, comparable to the increase in the length
 of the first time bin. Unfortunately, the relative uncertainty in $\Psi_2$ also becomes very large, precluding the
 possibility of saying with any confidence whether or not a SN~Ia component at delays of 80-330~Myr exists. However, it
 is highly unlikely that, at 80~Myr delays, there are still contributions to the DTD from CC-SNe, as this delay
 corresponds to the lifetime of stars with initial masses below $6M_\odot$. Thus, this exercise suggests that much of
 the signal in the 35-330~Myr bin of the DTD could be due to SNe~Ia that explode within $<80$~Myr. This result joins
 other evidence for prompt SNe~Ia with extremely short delays, of $\lesssim 100$~Myr
  \citep[; Della Valle et al. 2005]{mannucci06:two_progenitor_populations_SNIa}. The obstacle to obtaining a
 clearer answer, however, is the small number of SNRs. One needs to recall that, among the 77 SNRs in our sample, at
 most half and possibly only about 10\% are "driving" the SN~Ia signals in the second and third DTD bins, and hence the
 large uncertainty. A larger sample of SNRs could be obtained by combining SNRs from several nearby galaxies, as briefly
 discussed in \S8, below.

\begin{table*}
 \centering
 \begin{minipage}{150mm}
  \caption{\textit{Bona fide} CC-SNRs in the LMC}
  \begin{tabular*}{80mm}{@{}llc@{}}
    \hline
    SNR  &  Alternate  & CC Classification \\
    & Name & Criteria and Reference\footnote{EJ: Ejecta emission clearly indicative of a CC SN origin (usually, O-rich
      and Fe-poor); NS: SNR contains either a confirmed neutron star or strong evidence for a pulsar wind nebula.} \\
    \hline
    J0453.6$-$6829 & B0453$-$685 & NS; \citet{gaensler03:0453} \\
    J0505.9$-$6802 & N23 & NS; \citet{hughes06:N23} \\
    J0525.1$-$6938 & N132D & EJ; \citet{borkowski07:N132D} \\
    J0525.4$-$6559 & N49B & EJ; \citet{park03:N49B} \\	
    J0526.0$-$6604 & N49 & EJ\footnote{This SNR contains a magnetar, but the NS-SNR connection is disputed \citep[see discussion in][]{badenes09:SNRs_LMC}. Nevertheless, the ejecta composition leaves little doubt about the SN type.}; \citet{park03:N49} \\
    J0531.9$-$7100 & N206 & NS; \citet{williams05:SNR_N206} \\
    J0535.5$-$6916 & SNR1987A & historical CC-SN \\
    J0535.7$-$6602 & N63A & EJ; \citet{warren03:N63A} \\
    J0536.1$-$6734 & DEM L241 & NS; \citet{bamba06:DEM_L241} \\
    J0537.8$-$6910 & N157B & NS; \citet{chen06:N157B} \\
    J0540.2$-$6919 & B0540-693 & NS; \citet{kirshner89:0540} \\
    \hline 
  \end{tabular*}
\end{minipage}
\end{table*}

\subsection{Comparison of the visibility time to theoretical
    estimates}
\label{compvistime}
An interesting question that we can ask, at this point, is how do the
observed values of visibility time that we have obtained compare to the
theoretical expectations for the age at the SNR transition to the radiative
phase. Blondin et al. (1998) found, using 1D hydrodynamical simulations
and a full cooling curve (rather than a power-law approximation) that
the transition time is well approximated by $1.6~t_{\rm cool}$ for a wide
range of conditions, where $t_{\rm cool}$ is their rough estimate for
the cooling time, as in our
Eqns.~\ref{eqtcool} and \ref{tmaxwithepsilon}, assuming a cooling function
represented by an
$\epsilon=-1$ power law. Specifically, they found that for an ambient
hydrogen number density of $1~{\rm cm}^{-3}$, and for the fiducial explosion
energy and gas abundances, $t_{\rm cool}\approx 29$~kyr, and hence
the theoretical transition time for this density is $t_{\rm max}\approx
29\times 1.6=46$~kyr.

Unfortunately, we cannot directly compare this number to the visibility
times we have obtained in this work, using the various density tracers.
Our tracers are all based on column densities, rather than
densities, under the assumption that the two are correlated.
For example, although we know that for the mean HI column density
of the clouds, $1.5\times 10^{21} {\rm cm}^{-2}$, the visibility time is
13.3~kyr (using HI as a tracer), we do not know what absolute density
this corresponds to. 
Therefore, neither do we know
how to scale the visibility time to its value for the fiducial Blondin et
al. (1998) density of $1~{\rm cm}^{-3}$.

Instead, we will check if the LMC disk
thickness needed in order to match the numbers is reasonable. Assuming a
visibility-time dependence on density of $t_{\rm max}\propto \rho^{-1/2}$,
our observed range of $t_{\rm max}=13.3$ to 22.5~kyr implies a mean density of
$n\sim[46/(13.3~{\rm to}~22.5)]^2\approx 4~{\rm to}~12 ~{\rm cm}^{-3}$. We note that, although Blondin
et al. (1998) consider the case of a shock expanding into a neutral
hydrogen ambient medium, the results would change little if the
ambient medium consisted of molecular hydrogen, or a mixture of neutral
and molecular hydrogen. The 21~cm emission traces only the atomic
hydrogen, while molecular hydrogen is present in the gas disk as well.
Norikazu et al. (2001) have used CO as a molecular gas tracer to 
measure a molecular-to-atomic mass ratio of
$0.2 : 1$
in the low-radial velocity component of the gas in the LMC disk, associated
with the densest regions. However, for low columns or low
metallicities, CO may no longer be a be a good tracer, because it can
be photodissociated whereas H$_2$ remains self-shielded (Leroy et
al. 2009), and the actual molecular ratio is likely higher.
Thus, it is reasonable to assume that the
total gas column density is at least $\approx 2\times 10^{21} {\rm cm}^{-2}$.
Dividing by $n$ then gives a LMC gas disk thickness of about 60 to
170~pc or more,
smaller than, but comparable to the 300~pc-thick Milky Way gas disk. We conclude
that, while there is too much missing information about the LMC geometry
(and certainly about that of the SMC) for a definitive test of the Blondin
et
al. (1998) model and its application by us to SNR visibility, there
appears to be rough consistency in the numbers.

\section{The Current SN Rate}
We have focused, up to this point, on deriving the SN DTD, which has permitted us
to separate, based on delay times, the contributions of CC-SNe, 
prompt SN~Ia, and delayed SNe~Ia. However, the data allow deriving
also the more traditional {\it current} SN rate in the MCs, which is also
of interest. Naturally, the rates we will get will be total rates, for
all types of SNe, CC and type-Ia, that produce SNRs. The SN rate in the LMC or SMC, or 
in both galaxies treated as a whole, will simply be
\begin{equation}
R=\frac{N_{\rm SNR}N_{\rm cells}}{\sum_i t_i},
\end{equation}
where $N_{\rm SNR}$ is the total number of remnants, $N_{\rm cells}$
is the number of cells, and the
sum over the visibility times, $t_i$, is over all cells. Alternatively,
one can calculate the SN rate per unit mass,
\begin{equation}
R_M=\frac{N_{\rm SNR}}{\sum_{i} m_i t_i},
\end{equation}
 where $m_i$ is the time-integrated stellar mass formed in each cell
 (i.e., $m_i=\sum_{j} m_{ij}$). 
The relative error in these rates is given by the sum in quadrature
of the relative Poisson error, $N_{\rm SNR}^{-1/2}$, and the relative
error of the visibility time. Since all the visibility times have been
 renormalised to give $\Psi_1=2.86$, the relative visibility-time error is just the relative
error in $\Psi_1$. 
Measurements of $R_M$ in other galaxies
have generally been normalised not to the formed stellar mass, but to
the existing stellar mass. For ease of comparison to those measurements, we
 will therefore scale down all values of $m_i$ by 0.7, which accounts
roughly for the mass loss of stars during stellar evolution in
 actively star-forming galaxies like the MCs
 \citep{bruzual93:isochrones}. (This 30\% mass loss is in contrast to the
50\% mass loss previously considered for old and quiescent stellar populations).
 Rates per unit
 mass
in Table~2 are in units of
SNuM [SNe~$(100~{\rm yr})^{-1}(10^{10}M_\odot)^{-1}$]. 

As with our derivation of the DTD,
we will get different rates, depending on the density tracer we use.
 Table~2 lists, for each of the density tracers, and for the hybrid
 combinations,
the values or $R$ 
and $R_M$ in the MCs as a whole. Considering the range in $R$ covered
by the uncertainties and by the different tracers, we see that a SN
explodes in the MCs once every 200 to 500 years. This is in excellent
agreement with the historical record: the two most recent SNe in the
MCs
were SN1987A (a CC-SN), and 
SNR J0509.5$-$6731
(B0509$-$67.5) 
which exploded {\it circa} 1600 and was a SN~Ia 
[\citep{rest08:0509}; see \S5.3 of \citep{badenes08:0509}].
It is highly implausible that many additional SNe exploded in the
MCs over the past 400 years but were unnoticed by naked-eye observers.
 
In terms of mass-normalised SN rates, the range encompassing the 
different tracers and their $1\sigma$ uncertaities, 1.7 to 3.7 SNuM,
 is in
excellent agreement with rates measured in the bluest dwarf
galaxies, of which the MCs are prototypes. For such galaxies,
which have the highest-known mass-normalised SN rates, 
\citep{mannucci05:SN_rate} have measured rates, in SNuM, of
$2.3\pm 0.8$ (CC-SNe) and $0.9\pm0.4$ (SNe~Ia), or a total SN rate
of $3.2\pm0.9$~SNuM. 

Conversely, we can use the above SN rates to reinforce our claim that
that SNR sample we compiled in Paper~I is indeed fairly
complete, and that our model, in which the MC SNRs are in their Sedov-Taylor phase, is correct. Suppose, by way of
negation, that the MC SNRs are actually in free expansion, as has been invoked by a number of authors to explain the
observed uniform SNR size distribution (see Paper~I). Since free-expansion SNR shock velocities are in the range $\sim
5000$ to 10,000~km~s$^{-1}$, the maximum observed SNR radii of $r\sim 30$~pc would translate to maximum ages of 3 to 6
kyr. The 77 SNRs in the MCs would then imply a SN explosion in the MCs once every 40 to 80 years, on average, or even
more frequently if our SNR sample is incomplete.  Similarly, the 3 to 6~kyr visibility time, independent of ambient
density in this picture, would give mass normalised total SN rates of 8 to 15~SNuM, or even higher if our sample is
incomplete. This is much higher than in any known type of galaxy. Most seriously, perhaps, if the visibility time were
fixed at, say, 6~kyr, this would raise $\Psi_1$ in the ``no scaling'' model of Table~2 from a value of 2.86 to $6.4~{\rm
  SNe~yr^{-1}}~(10^{10}~M_{\odot})^{-1}$, corresponding to a CC-SN yield of $0.022~M_\odot^{-1}$. To produce so many
CC-SNe with any standard IMF, all stars with initial mass above $3.7 M_\odot$ would have to undergo core-collapse. This
is in direct contradiction to stellar evolution theory, and to the semi-empirical initial-final mass relation for WDs
(e.g. \citep{catalan08:initial_final_mass_WDs,salaris09:WD_initial_mass_function,williams09:lower_mass_SN_progenitors})

In summary, the Sedov model we have adopted to explain the MC SNR size distribution is consistent with the gas density
distributions of the MCs (Paper~I), and results in MC SN rates that are consistent with the historical record and with
the rates measured in similar types of galaxies. In contrast, a free-expansion model for the MC SNRs produces rates that
are in stark contradiction with these observations, and with expectations from stellar evolution.

\section{Summary and Conclusions}
In Paper~I, we assembled a multi-wavelength compilation of the 77 known SNRs in the MCs, collected from the existing
literature.  We verified that this compilation is fairly complete, and that the size distribution of SNRs is
approximately flat, within the allowed uncertainties, up to a cutoff at $r\sim30$ pc, as noted by other authors before.
We then proposed a physical model to explain this size distribution. According to our model, most of the SNRs are in
their Sedov stage, quickly fading below detection as soon as they reach the radiative stage. Under these circumstances,
a flat distribution of SNR sizes can be obtained, provided that the distribution of densities in the MCs follows a power
law with index $-1$.  Finally, in Paper~I, we used three different density tracers (HI column density, H$\alpha$ flux,
and SFR) to demonstrate that the distribution of densities in the MCs indeed follows a power law of index $-1$.

In this paper, we have used the Paper~I sample of SNRs as an effective supernova survey, conducted
over tens of kyr. 
We have applied a novel technique to this SNR sample to derive,
in these galaxies,
the SN rates and the SN DTD. In order to accomplish
this, we have
used the three tracers of the density to scale the visibility time of SNRs in the MCs. From the scaled visibility times
and the SFH maps from 
the resolved stellar populations in the MCs published by
  \cite{harris04:SMC_SFH} and \cite{harris09:LMC_SFH}, we have
  calculated the DTD. 
The DTDs we have derived are the first obtained using SFHs from resolved stellar populations.
  
Our main findings are the following:\\
1. We detect at the $>99\%$ confidence level a ``prompt'' SN~Ia
population, defined here as one that explodes within 330~Myr of star
formation. This finding joins a growing number of measurements of such
a component (see \S1). However, our measurement is based on our
statistically robust DTD recovery method, with its avoidance of the
averaging inherent to many previous measurements, and using the most
reliable SFHs, based on resolved stellar populations.
The yield of prompt SN~Ia, in terms of SNe per stellar mass
formed, is also consistent with other measurements.
\\
2. We obtain upper limits on the delayed SN~Ia population, which are
consistent with other measurements. Using these upper limits, we find
that roughly half, but possibly more, of  SNe~Ia are ``prompt'', as defined above. This
again joins previous results on the large relative fraction of
the prompt population.\\
3. We use our SNR sample and our SNR visibility times to derive
current MC SN rates. The SN rate in the MCs agrees well with the
historical record for these galaxies. The mass-normalised SN rate
in the MCs is in excellent agreement with the rates measured by SN
surveys in galaxies of this type. This lends support to the physical
model we have presented to explain the SNR size distribution and the
clouds, and which we have used to derive visibility times for our DTD
and rate calculations.\\
4. Conversely, a ``free expansion'' model for the SNRs, as has been 
invoked for the MCs and other galaxies, would imply SN rates in 
conflict with the historical record and with the rates in other
star-forming dwarf galaxies. Furthermore, this model would indicate an
unreasonably high yield of CC-SNe per stellar mass formed.

The main limitation of our study has been the relatively small number
of SNRs in the MC sample, which has forced us to use coarse time bins
and has led to large Poisson
errors. Construction of significantly larger samples of SNRs is, however,
possible by means of deep radio surveys of additional galaxies that
are near enough for deriving SFHs of their individual regions via
resolved stellar populations, namely M31 and M33 (see Paper I). 
Such data would permit a similar
analysis
to the one we have done, but with larger SNR numbers,
permitting a more accurate determination
of the SN DTD, and bringing into
better focus the properties of SNe, their remnants, and the connections
between them.

\section*{Acknowledgments}
We thank Dennis Zaritsky for many helpful discussions about several details of the SFH maps of the MCs. We acknowledge
useful interactions with Bryan Gaensler, Avishay Gal-Yam, Jack Hughes,
Jose Luis Prieto, Amiel Sternberg, Jacco Vink, and Eli Waxman, and
useful comments by the anonymous referee.
D.M. acknowledges support by the Israel Science Foundation and by the DFG through German-Israeli Project Cooperation
grant STE1869/1-1.GE625/15-1. C.B. thanks the Benoziyo Center for Astrophysics for support at the Weizmann Institute of
Science. This research has made use of NASA's Astrophysics Data System (ADS) Bibliographic Services, as well as the
NASA/IPAC Extragalactic Database (NED).

\bibliographystyle{mn2e}

\end{document}